\documentclass[conference, a4paper]{IEEEtran}
\IEEEoverridecommandlockouts

\usepackage{cite}
\usepackage{graphicx}
\usepackage{amsmath}
\usepackage{array}
\usepackage{mdwmath}
\usepackage{amssymb}
\usepackage{mdwtab}
\usepackage{stfloats}
\usepackage{subcaption} % 新版，功能更全，若用则不要再加载 subfig
\usepackage{amsmath,amsthm}
\usepackage{threeparttable}
\usepackage{color}
\usepackage{url}
\usepackage{algpseudocode}
\usepackage{algorithm}
\usepackage{multicol}
\usepackage{amssymb}
\usepackage{epstopdf}
\usepackage{flushend}
\algrenewcommand{\algorithmicrequire}{\textbf{Input:}}
\algrenewcommand{\algorithmicensure}{\textbf{Output:}}

\newtheorem{lemma}{Lemma}

\def\BibTeX{{\rm B\kern-.05em{\sc i\kern-.025em b}\kern-.08em
    T\kern-.1667em\lower.7ex\hbox{E}\kern-.125emX}}

\begin{document}
\title{On the Fundamental Scaling Laws of Fluid Antenna Systems}

\author{
    \IEEEauthorblockN{
      Xusheng Zhu$^1$, Farshad Rostami Ghadi$^1$, Tuo Wu$^2$, Kaitao Meng$^3$, Chao Wang$^4$, Gui Zhou$^5$
}
    \IEEEauthorblockA{1: Department of Electronic and Electrical Engineering, University College London, London, UK}
    \IEEEauthorblockA{2: Department of Electronic
Engineering, City University of Hong Kong, Hong Kong }
    \IEEEauthorblockA{3: Department of Electrical and Electronic Engineering,  University of Manchester, UK }
    \IEEEauthorblockA{4: State Key Laboratory of Integrated Service Network, Xidian University, Xi'an 710071, China}
    \IEEEauthorblockA{5: School of Electronic Information and Communication, Huazhong University of Science and Technology (HUST), China}
    %\IEEEauthorblockA{Corresponding Author: Chao Wang \quad Email: drchaowang@126.com}
    \thanks{Corresponding Author: Chao Wang (Email: drchaowang@126.com). This work was supported in part by the Engineering and Physical Sciences Research Council (EPSRC) under Grant EP/W026813/1; in part by
NSFC Grant 624B2094; in part by the Outstanding Doctoral
Graduates Development Scholarship of Shanghai Jiao Tong
University.}
}

\markboth{}{}

\maketitle
\begin{abstract}

Fluid antenna systems (FAS) offer a promising paradigm for enhancing wireless communication by exploiting spatial diversity, yet a rigorous analytical framework for their error probability has been notably absent. To this end, this paper addresses this critical gap by unveiling the \textbf{fundamental scaling laws} that govern the symbol error rate (SER) of FAS in realistic, spatially correlated channels. To establish these laws, we derive a tight, closed-form asymptotic expression for the SER applicable to a general class of modulation schemes. 
This result is pivotal as it establishes the fundamental scaling law governing the relationship between SER and the channel's spatial correlation structure. Based on this framework, we provide a complete characterization of the diversity and coding gains. 
The analysis culminates in a definitive design directive: SER can be fundamentally improved by expanding the antenna's movement space to increase diversity, while merely increasing port density within a constrained space yields diminishing returns.
\end{abstract}


\section{Introduction}
Modern wireless communications rely on multiple-input multiple-output (MIMO) systems to enhance reliability and capacity \cite{wong2022extrm, new2024an, zhu2025disc,ris2025z}. However, their static antenna configuration is a key limitation, as performance degrades when an antenna is stuck in a deep fade\cite{zhou2024movab}. 

% The fluid antenna system (FAS) paradigm was introduced to overcome this by enabling an antenna to dynamically move within a defined area to find the best signal conditions \cite{wong2022bruce, new2025at, wong2021flns, wu2024flu}. By selecting the port with the highest instantaneous signal-to-noise ratio (SNR), FAS achieves significant diversity gains without the constraints of conventional fixed arrays.
% Prior research has extensively explored FAS performance, with foundational tutorials establishing its basic principles \cite{wong2020fles, wong2023ii, wong2023III}. Performance analysis has largely centered on outage probability and ergodic capacity \cite{cha2024onms, wong2020pems}, and has extended the concept to multi-user scenarios through fluid antenna multiple access (FAMA) \cite{kwonglimi202, wong2023oppf, won2023trsm}. 

To this end, the fluid antenna system (FAS) was developed to address the limitations of traditional static antenna systems 	\cite{wong2022bruce,wu2025scalable,yao2025fas}. By allowing dynamic antenna positioning, FAS enables small movements of the antenna to significantly enhance signal strength\cite{hong2025ge}. A single antenna can switch between multiple closely spaced locations within a defined area, allowing the system to move from weak signal zones to stronger ones with minimal displacement \cite{new2025at, wong2021flns, wu2024flu}. This dynamic feature of FAS ensures the receiver can always select the port with the best current signal, providing reliable and robust communication under changing conditions. To facilitate FAS adoption, a comprehensive three-part tutorial series has been published \cite{wong2020fles, wong2023ii, wong2023III}, covering fundamental principles, research opportunities, and emerging applications, thereby laying a solid foundation for further exploration. FAS considers antenna positions as discrete ports, with the system selecting the port with the highest instantaneous signal-to-noise ratio (SNR) for reception. This ability to dynamically choose the best port effectively provides multi-antenna gains at the receiver without the constraints of physical antenna arrays.

The performance of FAS has been extensively analyzed. For instance, \cite{cha2024onms} demonstrates that as the number of available ports increases, the outage probability decreases significantly, with FAS outperforming maximal-ratio combining (MRC) when the port count is sufficiently high. Building on this, \cite{wong2020pems} derives second-order statistics, such as ergodic capacity and average fade duration, which provide deeper insights into the long-term behavior of FAS in various channel conditions. In addition, the extension of FAS to multi-user environments, through fluid antenna multiple access (FAMA), is explored in \cite{kwonglimi202}, where it is shown that a single fluid antenna can support hundreds of users simultaneously, making FAS a scalable solution for next-generation wireless networks. Furthermore, \cite{won2023trsm} introduces the compact ultra-massive antenna (CUMA) receiver architecture, designed to protect the desired user's signal from interference, even in the absence of channel state information (CSI) from interferers. This architecture leverages the adaptability of fluid antennas to enhance signal quality and reduce interference in dynamic environments.

Despite extensive analysis of FAS focusing on metrics like outage and capacity, a significant gap persists in understanding its error performance. symbol error rate (SER)\footnote{When the modulation order equals 2, the SER is degenerated to the bit error rate (BER).}, a critical metric for link-level design, remains largely uncharacterized within a rigorous analytical framework. This absence hinders accurate performance evaluation and the formulation of effective design principles. This paper aims to bridge this critical gap by providing the first comprehensive, SER-based performance analysis of FAS.
The main contributions of this paper are summarized as follows:
1) We derive a tight, closed-form asymptotic expression for the SER of FAS, thereby establishing the fundamental scaling law that governs its performance. This result provides a powerful tool for precise performance prediction and reveals the intrinsic relationship between SER and the channel's spatial correlation structure.
2) We provide a complete characterization of the system's diversity and coding gains based on this scaling law. Our analysis rigorously decouples the impact of the physical channel structure from the signal constellation design.
3) We formulate a definitive design directive derived from our analysis: expanding the antenna's movement space is the primary method for performance enhancement, while merely increasing port density within a fixed area yields diminishing returns.

% The main contributions of this paper are summarized as follows:
% 1) We establish the \textbf{fundamental scaling law} that governs the SER of FAS by deriving a tight, closed-form asymptotic expression. This result provides a powerful tool for precise performance prediction and reveals the intrinsic relationship between SER and the channel's spatial correlation structure.
% 2) Based on this scaling law, we provide a complete characterization of the system's \textbf{diversity and coding gains}. Our analysis culminates in a definitive design directive: expanding the antenna's movement space is the primary method for performance enhancement, while merely increasing port density within a fixed area yields diminishing returns.

\begin{figure}[t]
  \centering
  \includegraphics[width=8cm]{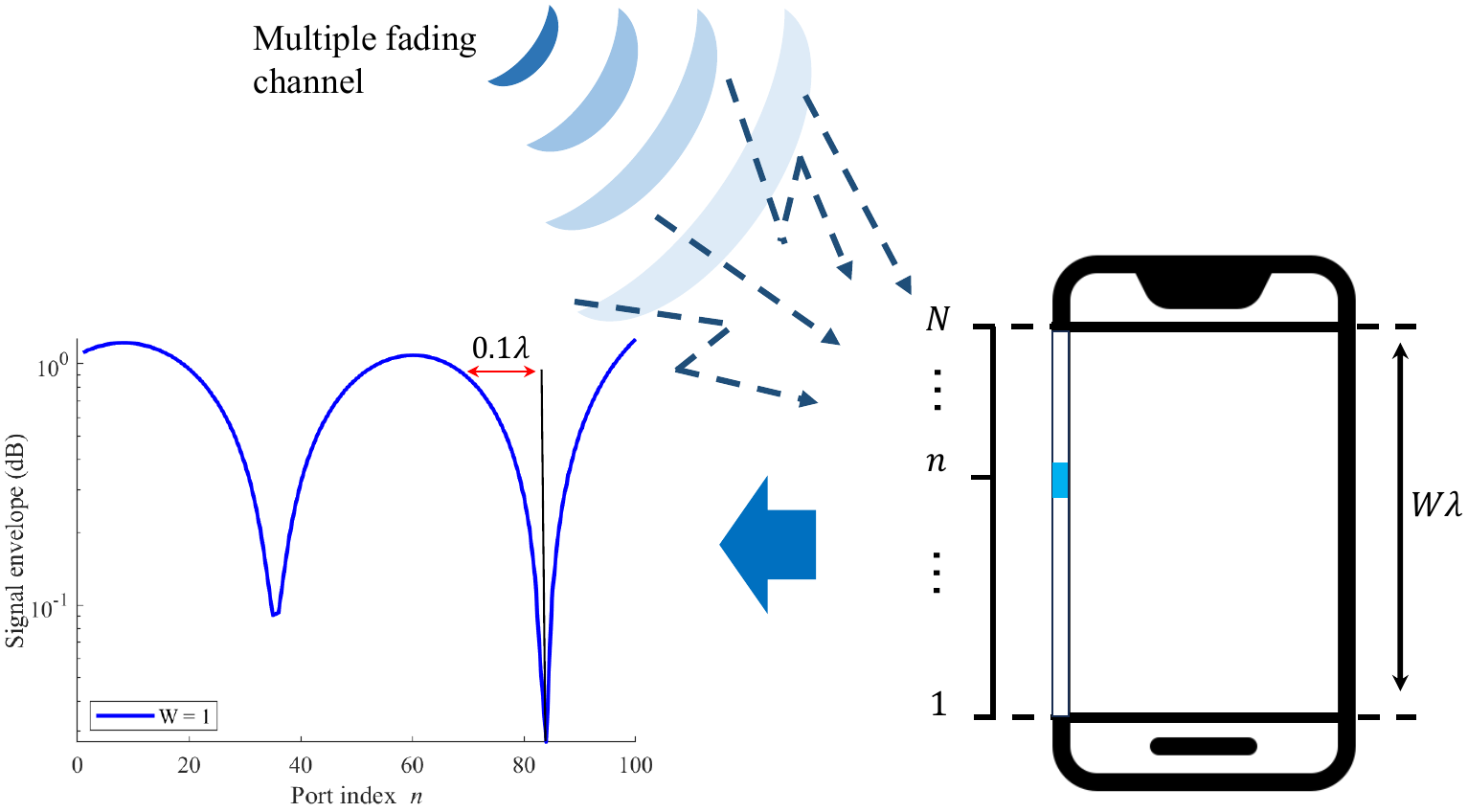}
  \caption{System model.}\label{framew}
\end{figure}

\textit{Notations}: 
The operators $(\cdot)^T$ and $(\cdot)^H$ denote transpose and Hermitian transpose. $\mathrm{E}[\cdot]$ is the statistical expectation. $|\cdot|$ is the absolute value or magnitude, and $\|\cdot\|$ is the Euclidean norm. $\mathrm{diag}(\cdot)$ creates a diagonal matrix. $\det(\cdot)$ and $\mathrm{Rank}\{\cdot\}$ denote the determinant and rank. $(\cdot)!!$ is the double factorial. $J_0(\cdot)$ is the zeroth-order Bessel function, $Q(\cdot)$ is the Gaussian Q-function, and $\Gamma(\cdot)$ is the Gamma function. $\mathcal{CN}(\mu, \sigma^2)$ denotes a circularly symmetric complex Gaussian distribution with mean $\mu$ and variance $\sigma^2$. $\mathbf{I}_N$ is the $N \times N$ identity matrix.

\section{System Model}
We consider a fundamental receive FAS model, as depicted in Fig. \ref{framew}, where a single antenna element can access any of $N$ uniformly spaced ports. These ports are distributed along a linear aperture of length $W\lambda$, with $\lambda$ being the carrier wavelength. The spatial separation between the $i$-th and $j$-th ports is defined as\footnote{This spatial formulation is fundamental to the subsequent analysis of spatially correlated fading, where the correlation structure is critically dependent on the inter-port distances.}
\begin{equation}\label{Deltadij}
\Delta d_{i,j} = \frac{|i - j|}{N - 1}W\lambda, \quad \text{for } i,j \in \{1, 2, \ldots, N\},
\end{equation}
where $W$ is the normalized aperture width.
The signal received at the $n$-th port is modeled as
\begin{equation}\label{eq2}
y_n = h_n s + w_n= g_n e^{j\phi_n} s + w_n, \quad n \in \{1, 2, \ldots, N\},
\end{equation}
where $s$ is the transmitted symbol, drawn from an $M$-ary constellation, e.g., phase-shift keying (PSK), quadrature amplitude modulation (QAM), or pulse amplitude modulation (PAM). $w_n \sim \mathcal{CN}(0, 1)$ is the additive white Gaussian noise (AWGN). The system employs a port selection strategy, choosing the port with the maximum instantaneous channel gain $g_{\rm FAS} = \max\limits_{n\in\{1,\cdots,N\}} \{g_n\}$. Consequently, the instantaneous SNR is given by
\begin{equation}\label{seglabe}
\gamma_{\rm FAS} = \bar{\gamma} |g_{\rm FAS}|^2,
\end{equation}
where $\bar{\gamma}$ denotes the average SNR per port.

The system's performance is fundamentally governed by the spatial correlation between ports. We model the channel coefficients as a vector $\mathbf{h} = [h_1, \ldots, h_N]^T$, whose statistical properties are captured by the covariance matrix $\mathbf{R} = \mathrm{E}[\mathbf{h}\mathbf{h}^H]$. This matrix can be expressed as $\mathbf{R} = \mathbf{T} \mathbf{J} \mathbf{T}$, where $\mathbf{T} = \mathrm{diag}(\sqrt{\bar{\gamma}_1}, \sqrt{\bar{\gamma}_2}, \ldots, \sqrt{\bar{\gamma}_N}) $ is a diagonal matrix of root-mean-square signal levels and $\mathbf{J}$ is the normalized spatial correlation matrix. Adopting the widely used Jake's model, the entries of $\mathbf{J}$ are given by $J_{i,j} = J_0\left(2\pi W \frac{|i-j|}{N-1}\right)$.
To facilitate the analysis, we perform an eigenvalue decomposition of the Hermitian matrix $\mathbf{J}$ as $\mathbf{J} = \mathbf{U} \boldsymbol{\Lambda} \mathbf{U}^H$, where $\mathbf{U}$ is a unitary matrix of eigenvectors and $\boldsymbol{\Lambda} = \mathrm{diag}(\lambda_1, \ldots, \lambda_N)$ is a diagonal matrix of corresponding real, non-negative eigenvalues, sorted in descending order. This allows the correlated channel vector to be expressed as a linear transformation of i.i.d. variables:
\begin{equation}
    \mathbf{h} = \mathbf{U}\boldsymbol{\Lambda}^{\frac{1}{2}} \mathbf{z},
\end{equation}
where $\mathbf{z} \sim \mathcal{CN}(\mathbf{0}, \mathbf{I}_N)$. This representation is instrumental for the performance analysis in the subsequent sections.

% \begin{table}[t] % [t] 表示优先将表格放在页面顶部
% \centering % 居中显示
% \caption{Parameters $p$ and $k$ for  Modulation Schemes}
% \label{tab:modulation_table}
% % 使用 resizebox 命令将 tabular 环境缩放到与列宽一致
% \resizebox{\linewidth}{!}{%
%     \begin{tabular}{|c|c|c|c|}
%     \hline
%     \textbf{Modulation } & \textbf{Conditional SER} & \textbf{$p$} & \textbf{$k$} \\
%     \hline
%     BPSK & $Q\left(\sqrt{2\gamma}\right)$ & $1$ & $2$ \\
%     \hline
%     $M$-PSK, $M \geq 4$ & $\approx 2Q\left(\sqrt{2\gamma} \sin\frac{\pi}{M}\right)$ & $2$ & $2\sin^2\left( \frac{\pi}{M} \right)$ \\
%     \hline
%     $M$-PAM & $2\left(1 - \frac{1}{M}\right) Q\left(\sqrt{ \frac{6\gamma}{M^2 - 1} }\right)$ & $2\left(1 - \frac{1}{M}\right)$ & $\dfrac{6}{M^2 - 1}$ \\
%     \hline
%     $M$-QAM & $4\left(1 - \frac{1}{\sqrt{M}}\right) Q\left(\sqrt{ \frac{3\gamma}{M - 1} }\right)$ & $4\left(1 - \frac{1}{\sqrt{M}}\right)$ & $\dfrac{3}{M - 1}$ \\
%     \hline
%     \end{tabular}%
% }
% \end{table}

\section{Performance Analysis}
In this section, we derive a closed-form asymptotic expression for the average SER of the FAS. By leveraging high-SNR asymptotic techniques, we obtain a tractable SER expression that provides fundamental insights into the FAS.

\subsection{Conditional Error Probability}
We first consider a coherent modulation with conditional SER as \cite{wang2003a}
\begin{equation}\label{cpep}
P_e(x)=pQ(\sqrt{kx\bar{\gamma}}),
\end{equation}
where $p$ and $k$ are constants related to the modulation format. 
The modulation-dependent parameters $p$ and $k$ are defined as follows: for BPSK, $p=1$ and $k=2$; for $M$-PSK ($M \geq 4$), $p=2$ and $k=2\sin^2(\pi/M)$; for $M$-PAM, $p=2(1 - 1/M)$ and $k=6/(M^2 - 1)$; and for $M$-QAM, $p=4(1 - 1/\sqrt{M})$ and $k=3/(M - 1)$.

% Because \eqref{cpep} contains the Q-function in its integral form, a closed-form expression is not attainable. This limitation obscures analytical insight and prevents a precise assessment of how each system parameter influences performance.

\subsection{Unconditional Error Probability}
The average SER is fundamentally defined as the expectation of the instantaneous SER over all possible channel realizations characterized by their probability density function (PDF). Mathematically, the instantaneous SER can be expressed as
\begin{equation}\label{PExpn}
\begin{aligned}
P_{E} = p \int_{0}^{\infty} f(x)Q\left(\sqrt{{kx\overline{\gamma}}}\right) dx = p\bar{P}_e,
\end{aligned}
\end{equation}
where $\bar{P}_e = \int_{0}^{\infty} f(x)Q\left(\sqrt{kx\overline{\gamma}}\right) dx$ denotes the average SER contribution from the PDF of the random variable $x$.

To proceed, we employ Lemma~1 to obtain the high-SNR approximation of $f(x)$.
\begin{lemma}
In the high-SNR regime, the PDF of $x=\gamma_F$ can be approximated as
\begin{equation}
f(x) \approx \frac{N x^{N-1}}{\det(\mathbf{J}) \prod_{n=1}^N \bar{\gamma}_n}, \quad \text{as } x \to 0^+.
\end{equation}
\end{lemma}

\emph{Proof:} See Appendix~A. \hfill$\blacksquare$

Substituting the result of Lemma~1 into the average SER expression yields the asymptotic form
\begin{equation}
\bar{P}_{e} \approx \int_{0}^{\infty} \frac{N x^{N-1}}{\det(\mathbf{J}) \prod_{n=1}^N \bar{\gamma}_n} Q\left(\sqrt{kx\overline{\gamma}}\right) dx.
\end{equation}
Using the integral representation of the $Q$-function\cite{zhu2025toward,zhu2024on},
$Q(z) = \frac{1}{\sqrt{2\pi}} \int_z^\infty e^{-v^2/2} dv$,
we can rewrite the expression as
\begin{equation}
\bar{P}_e \approx \frac{N}{\det(\mathbf{J}) \prod_{n=1}^N \bar{\gamma}_n \sqrt{2\pi}} \int_{0}^{\infty} x^{N-1} \int_{\sqrt{kx\overline{\gamma}}}^{\infty} e^{-v^2/2} dv  dx.
\end{equation}
By interchanging the order of integration, where the domain is defined by $0 < x < \infty$ and $\sqrt{kx\overline{\gamma}} < v < \infty$, the integration limits are transformed to $0 < v < \infty$ and $0 < x < v^2/(k\overline{\gamma})$. The expression for $P_e$ becomes
\begin{equation}\label{barpeev}
\begin{aligned}
\bar P_{e} &\approx 
% \frac{N}{\det(\mathbf{J}) \prod_{n=1}^N \bar{\gamma}_n\sqrt{2\pi}} \int_{0}^{\infty} e^{-v^2/2} \left( \int_{0}^{\frac{v^2}{k\overline{\gamma}}} x^{N-1} dx \right) dv \\
% &= 
% \frac{N}{\det(\mathbf{J}) \prod_{n=1}^N \bar{\gamma}_n\sqrt{2\pi}} \int_{0}^{\infty} e^{-v^2/2} \left[ \frac{x^{N}}{N} \right]_0^{\frac{v^2}{k\overline{\gamma}}} dv \\
% &= \frac{N}{\det(\mathbf{J}) \prod_{n=1}^N \bar{\gamma}_n\sqrt{2\pi}} \int_{0}^{\infty} e^{-v^2/2} \left( \frac{1}{N} \left( \frac{v^2}{k\overline{\gamma}} \right)^{N} \right) dv \\
% &= 
\frac{1}{\det(\mathbf{J}) \prod_{n=1}^N \bar{\gamma}_n\sqrt{2\pi}(k\overline{\gamma})^{N}} \int_{0}^{\infty} v^{2N} e^{-v^2/2} dv.
\end{aligned}
\end{equation}
The remaining integral can be solved by relating it to the Gamma function
%\begin{equation}\label{gammazd}
$\Gamma(z) = \int_0^\infty y^{z-1}e^{-y}dy$.
%\end{equation}
Using the substitution $y = v^2/2$, which implies $v=\sqrt{2y}$ and $dv = (1/\sqrt{2y})dy$, we have
\begin{equation}
\begin{aligned}
\int_{0}^{\infty} v^{2N} e^{-v^2/2} dv 
% &= \int_{0}^{\infty} (\sqrt{2y})^{2N} e^{-y} \frac{1}{\sqrt{2y}} dy \\
% &= \int_{0}^{\infty} 2^N y^N e^{-y} 2^{-1/2} y^{-1/2} dy \\
% &= 2^{N-1/2} \int_{0}^{\infty} y^{N-1/2} e^{-y} dy \\
&= 2^{N-1/2} \Gamma\left(N+\frac{1}{2}\right),
\end{aligned}
\end{equation}
Accordingly, (\ref{barpeev}) can be reformulated as
\begin{equation}\label{barpeev1}
\begin{aligned}
\bar P_{e}
&\approx \frac{ 2^{N-1/2} \Gamma\left(N+\frac{1}{2}\right)}{\det(\mathbf{J}) \prod_{n=1}^N \bar{\gamma}_n\sqrt{2\pi}(k\overline{\gamma})^{N}}.
\end{aligned}
\end{equation}
By substituting this result into (\ref{PExpn}), we obtain 
\begin{equation}\label{Pepin}
\begin{aligned}
P_E &\approx
%\frac{p}{\det(\mathbf{J}) \prod_{n=1}^N \bar{\gamma}_n\sqrt{2\pi}(k\overline{\gamma})^{N}} \left( 2^{N-1/2} \Gamma\left(N+\frac{1}{2}\right) \right) \\
%&=
\frac{p2^{N-1} \Gamma(N+1/2)}{\det(\mathbf{J}) \prod_{n=1}^N \bar{\gamma}_n\sqrt{\pi}} (k\overline{\gamma})^{-N},
\end{aligned}
\end{equation}
where 
$\Gamma\left(N + \frac{1}{2}\right) = \frac{(2N - 1)!!}{2^N} \sqrt{\pi}$,
$(2N - 1)!! = (2N - 1)(2N - 3)\cdots(3)(1)$.
In this manner, we can obtain the asymptotic expression as
\begin{equation}\label{PEfinal}
P_E \approx \frac{p  2^{N - 1} }{\det(\mathbf{J}) \prod_{n=1}^N \bar{\gamma}_n \sqrt{\pi}}  \frac{(2N - 1)!!}{2^N} \sqrt{\pi}  (k\overline{\gamma})^{-N}.
\end{equation}
After some simplifications, we have
\begin{equation}\label{PEclosed}
P_E \approx \frac{p  (2N - 1)!!}{2 \det(\mathbf{J}) \prod_{n=1}^N \bar{\gamma}_n} (k\overline{\gamma})^{-N}.
\end{equation}
The derived asymptotic SER expression in (\ref{PEclosed}) reveals a fundamental \textbf{scaling law} for FAS performance. This law precisely dictates how the SER scales with two key factors: the average SNR and the channel's spatial correlation. The impact of spatial correlation is quantified by the determinant of the correlation matrix, $\det(\mathbf{J})$, which governs the scaling of the error probability floor. The behavior of this law is best understood by its extremes:
\begin{itemize}
    \item In the ideal case of uncorrelated ports, $\det(\mathbf{J}) \to 1$, which minimizes the SER and thus maximizes the diversity gain\footnote{We will provide a detailed derivation of the diversity gain of this system in the next subsection.}.
    \item Conversely, as ports become highly correlated, $\det(\mathbf{J}) \to 0$, causing the SER to approach infinity and effectively nullifying any diversity gain.
\end{itemize}
Therefore, this scaling law establishes a direct, quantitative link between the physical antenna geometry (which determines $\det(\mathbf{J})$) and the ultimate error performance of the system.

% The insights of the (\ref{PEclosed}) can be summarized as the Remak 1.
% \begin{remark}
% The asymptotic SER in (\ref{PEclosed}) provides an important analytical tool for rapidly and accurately estimating the error performance of FAS in the high-SNR regime.

% 1) The determinant of the correlation matrix, $\det(\mathbf{J})$, provides significant insight into the structure of the spatial channel and its impact on system performance. From matrix theory, the determinant is the product of its eigenvalues, $\prod_{n=1}^N \lambda_n$. For a normalized correlation matrix, the diagonal elements are unity, leading to the Hadamard inequality: $\det(\mathbf{J}) \leq \prod_{n=1}^N J_{n,n} = 1$. Equality holds if and only if $\mathbf{J}$ is a diagonal matrix, specifically $\mathbf{J} = \mathbf{I}_N$, which represents the ideal scenario of completely uncorrelated ports where spatial diversity is maximized.

% 2) Conversely, the other extreme, $\det(\mathbf{J}) = 0$, signifies a rank-deficient matrix. This condition implies linear dependence among the channel vectors at different port locations, which occurs when ports are so closely spaced within a limited aperture that their channel responses become highly correlated. In such a scenario, the system is unable to resolve distinct spatial paths, and the diversity gain from port switching is nullified.

% \end{remark}

\subsection{Diversity and Coding Gain Analysis}

In this subsection, we provide a rigorous analysis of the diversity and coding gains, which are two pivotal metrics that characterize the high-SNR performance of a communication system. 

\subsubsection{Diversity Gain Derivation}
The diversity gain, denoted by $G_d$, quantifies the rate at which the error probability decreases as the SNR increases. It is formally defined as the negative slope of the asymptotic SER curve when plotted on a log-log scale. Mathematically, this is expressed as
\begin{equation}
G_{d} = -\lim_{\overline{\gamma}\rightarrow\infty}\frac{\log P_{E}(\overline{\gamma})}{\log\overline{\gamma}},
\end{equation}
where $P_E(\overline{\gamma})$ is the average SER as a function of the average SNR, $\overline{\gamma}$.

% For correlated fading channels, the diversity order is determined by the number of independent spatial paths the channel can support, which corresponds directly to the effective rank of the channel correlation matrix, $N_{\rm eff} = \text{Rank}\{\mathbf{J}\}$. Therefore, to accurately model the system's behavior, the diversity order in the asymptotic SER expression from (\ref{PEclosed}) should be represented by $N_{\rm eff}$ instead of $N$. The adapted expression is
In correlated fading channels, the diversity order is not merely the number of ports $N$, but the number of independent spatial paths the channel can support. This corresponds directly to the effective rank of the channel correlation matrix, $N_{eff} = \text{Rank}\{J\}$. To accurately capture the system's behavior, we must therefore replace $N$ with $N_{eff}$ in our asymptotic SER expression from (16), yielding the adapted expression
\begin{equation}\label{reffg}
    P_E(\overline{\gamma}) \approx \left( \frac{p(2N_{\rm eff}-1)!!}{2k^{N_{\rm eff}} \det(\mathbf{J}) \prod_{n=1}^{N_{\rm eff}}\overline{\gamma}_{n}} \right) (\overline{\gamma})^{-N_{\rm eff}}.
\end{equation}
For high-SNR analysis, we can group the terms that do not depend on $\overline{\gamma}$ into a single constant, $C$. The expression then simplifies to the form $P_E(\overline{\gamma}) \approx C \cdot (\overline{\gamma})^{-N_{\rm eff}}$.

Substituting this simplified asymptotic SER into the definition of diversity gain, we have
\begin{equation}
G_d 
%= -\lim_{\overline{\gamma}\rightarrow\infty}\frac{\log(C \cdot \overline{\gamma}^{-N_{\rm eff}})}{\log\overline{\gamma}} 
= -\lim_{\overline{\gamma}\rightarrow\infty}\frac{\log C - N_{\rm eff}\log\overline{\gamma}}{\log\overline{\gamma}}.
\end{equation}
As $\overline{\gamma}\rightarrow\infty$, the term $\frac{\log C}{\log\overline{\gamma}}$ approaches zero. This leaves
\begin{equation}
G_d = -(-N_{\rm eff}) = N_{\rm eff} = \text{Rank}\{\mathbf{J}\},
\end{equation}
where the diversity gain of the FAS is equal to the effective rank of the channel correlation matrix.
This result highlights that the diversity performance is fundamentally governed by the channel's effective degrees of freedom, not merely the number of available ports $N$.

\subsubsection{Coding Gain Derivation}
The coding gain, $G_c$, represents the system's power efficiency and corresponds to a horizontal shift of the error curve. It can be derived by expressing the asymptotic SER in the canonical form $P_E \approx (G_c \overline{\gamma})^{-G_d}$. Using our result $G_d = N_{\rm eff}$, this becomes
\begin{equation}
P_E \approx (G_c \overline{\gamma})^{-N_{\rm eff}} = G_c^{-N_{\rm eff}} (\overline{\gamma})^{-N_{\rm eff}}.
\end{equation}
We now equate this with the detailed SER expression derived from (\ref{reffg}) where $\prod_{n=1}^{N_{\rm eff}}\lambda_n$, and we assume the normalized average SNR across ports to be
\begin{equation}
P_{E}\approx\left[ \frac{p(2N_{\rm eff}-1)!!}{2 \cdot k^{N_{\rm eff}} \cdot \prod_{n=1}^{N_{\rm eff}}\lambda_n} \right] (\overline{\gamma})^{-N_{\rm eff}}.
\end{equation}
By comparing the coefficients of the $(\overline{\gamma})^{-N_{\rm eff}}$ term in both SER expressions, we establish the following identity
\begin{equation}
G_c^{-N_{\rm eff}} = \frac{p(2N_{\rm eff}-1)!!}{2 \cdot k^{N_{\rm eff}} \prod_{n=1}^{N_{\rm eff}}\lambda_n}.
\end{equation}
To solve for $G_c$, we raise both sides of the equation to the power of $(-1/N_{\rm eff})$
% \begin{equation}
% G_c = \left( \frac{p(2N_{\rm eff}-1)!!}{2k^{N_{\rm eff}}} \right)^{-\frac{1}{N_{\rm eff}}} \cdot \left( \frac{1}{\prod_{n=1}^{N_{\rm eff}}\lambda_n} \right)^{-\frac{1}{N_{\rm eff}}}.
% \end{equation}
% This simplifies to the final expression for the coding gain
\begin{equation}
G_c = \left( \frac{2k^{N_{\rm eff}}}{p(2N_{\rm eff}-1)!!} \right)^{\frac{1}{N_{\rm eff}}} \left( \prod_{n=1}^{N_{\rm eff}}\lambda_n \right)^{\frac{1}{N_{\rm eff}}}.
\end{equation}
This rigorously derived result shows that the coding gain is determined by the modulation parameters ($p, k$) and the geometric mean of the effective channel gains, which are the non-zero eigenvalues of the correlation matrix. This provides a complete and accurate characterization of the high-SNR performance.

\section{Results and Discussion}
In this section, we present numerical results to validate our theoretical framework and offer further insights into the performance of FAS. modulation schemes.

\begin{figure}[t]
  \centering
  \includegraphics[width=7.0cm]{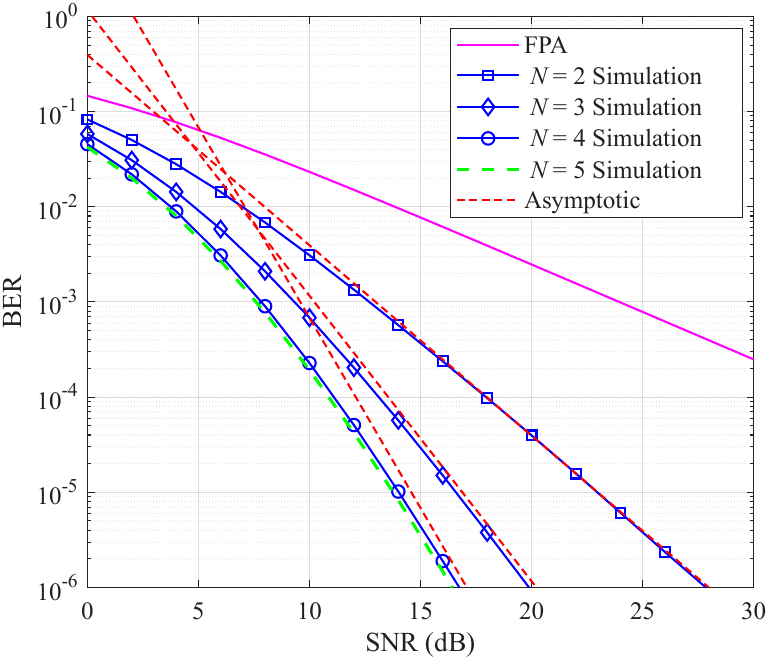}
  \caption{Validation of the asymptotic BER analysis for FAS with a normalized aperture ($W=1$).}
\label{fig:validation_N}
\end{figure}

Fig.~\ref{fig:validation_N}validates our asymptotic BER analysis for an FAS with a normalized aperture of $W=1$ using BPSK modulation. The results for $N \in \{2, 3, 4, 5\}$ show excellent agreement between the simulation results and the derived asymptotic expressions, particularly in the high-SNR regime.
Furthermore, the diversity gain is visually confirmed, as the slopes of the BER curves become progressively steeper with increasing $N$. This indicates a higher diversity order, with the slopes in the high-SNR region matching the theoretical gain of $G_d = N_{eff}$.
Even with just two ports ($N=2$), the FAS significantly outperforms a conventional fixed-port antenna (FPA), underscoring the fundamental benefit of exploiting spatial diversity.

\begin{figure}[t]
  \centering
  \includegraphics[width=7.0cm]{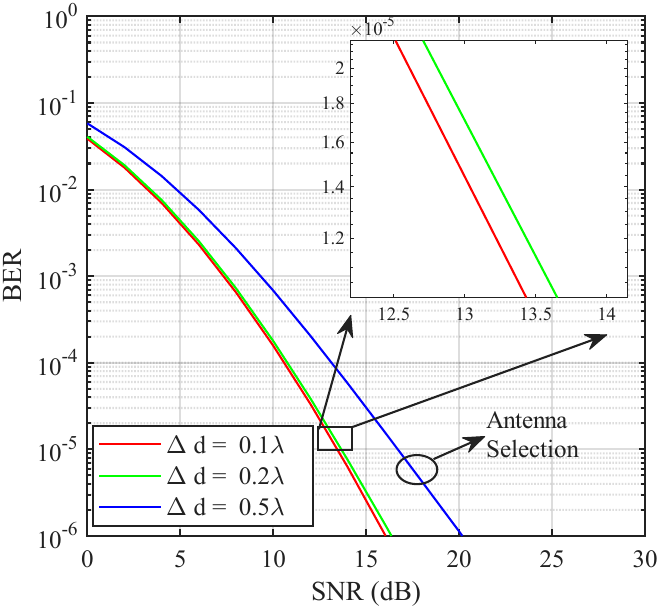}
  \caption{Impact of port spacing $\Delta d$ on the FAS performance for a fixed aperture ($W=1$).}
  \label{fig:port_spacing}
\end{figure}

Fig.~\ref{fig:port_spacing} investigates the impact of port density on system performance within a fixed normalized aperture of $W=1$. We compare three configurations with varying port spacing: $\Delta d = 0.5\lambda$ ($N=3$), $\Delta d = 0.2\lambda$ ($N=6$), and $\Delta d = 0.1\lambda$ ($N=11$). 
The results reveal that decreasing the port spacing (and thus increasing $N$ within a fixed aperture) yields diminishing returns in BER performance. While reducing the spacing from $0.5\lambda$ to $0.2\lambda$ provides a noticeable gain, the subsequent improvement from $0.2\lambda$ to $0.1\lambda$ is marginal.
Once the number of ports is sufficient to capture the essential spatial modes, adding more ports leads to higher spatial correlation and provides redundant channel information, resulting in negligible performance enhancement.

\begin{figure*}[!t]
    \centering
    \subfloat[PSK scheme.]{%
        \includegraphics[width=0.31\textwidth]{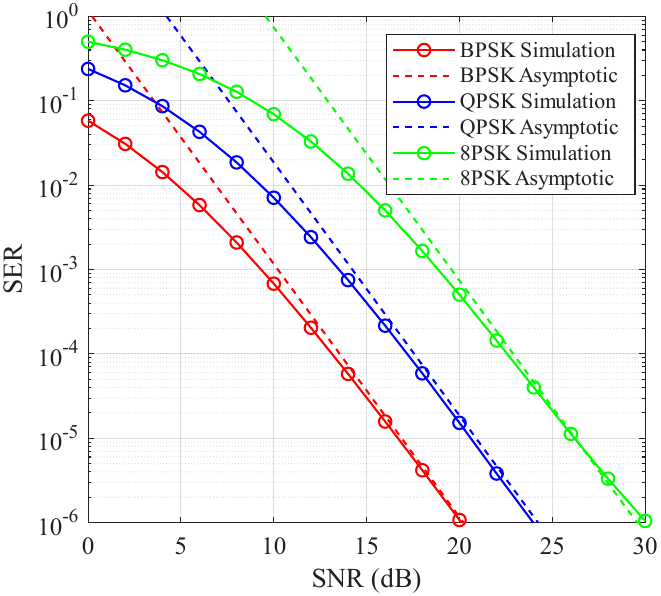}%
        \label{fig:psk}}
    \hfill
    \subfloat[QAM scheme.]{%
        \includegraphics[width=0.31\textwidth]{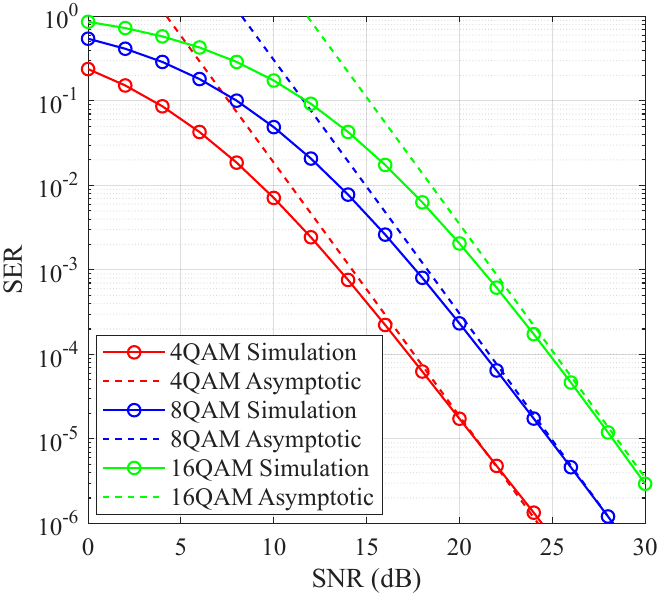}%
        \label{fig:qam}}
    \hfill
    \subfloat[PAM scheme.]{%
        \includegraphics[width=0.31\textwidth]{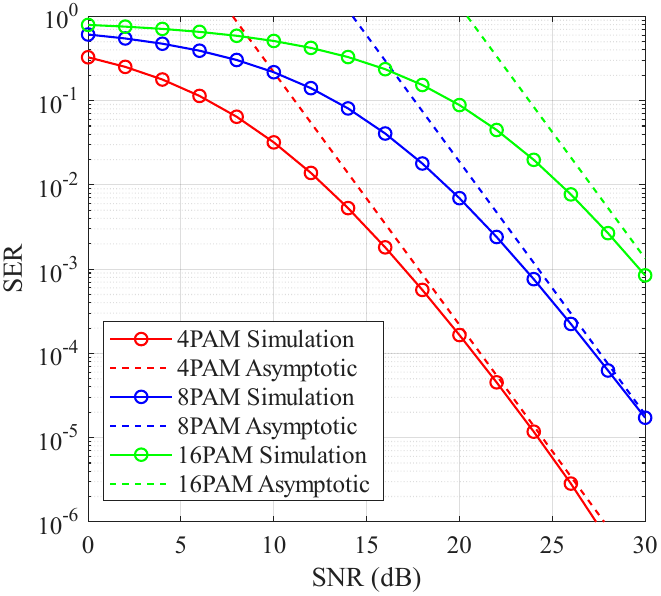}%
        \label{fig:pam}}
    \caption{SER performance comparison under various FAS modulation schemes ($N=3$, $W=1$). }
    \label{fig:modulation_compare}
\end{figure*}

Fig.~\ref{fig:modulation_compare} demonstrates the robustness and generality of our analytical framework by evaluating the SER performance for multiple coherent modulation schemes: PSK, QAM, and PAM. For this analysis, the FAS parameters are fixed at $N=3$ and $W=1$. Across all subplots and for all modulation orders, the simulation results show remarkable consistency with the high-SNR asymptotic expressions. This validates the generalized SER formula presented in (\ref{PExpn}).
A crucial insight from Fig. ~\ref{fig:modulation_compare} is that for a fixed physical system configuration ($N, W$), the SER curves for different modulation orders are parallel in the high-SNR region. This parallelism provides a compelling visual confirmation of a key theoretical finding: the diversity gain, dictated by the slope of the curve, is determined solely by the physical parameters of the FAS (via $\text{Rank}\{J\}$) and is therefore independent of the modulation format. The horizontal shifts observed between the curves are attributed to the differences in coding gain, which, as derived in our analysis, is a function of the modulation parameters $p$ and $k$. This result rigorously decouples the effects of the physical channel structure from the signal constellation design.

\section{Conclusions}
In this paper, we established the fundamental scaling laws governing the error performance of FAS. By deriving a tight, closed-form asymptotic expression for the SER, we have provided a robust and precise analytical framework, whose accuracy was rigorously validated against simulations across various system configurations and modulation schemes. Our analysis delivered a complete characterization of diversity and coding gains, successfully decoupling the effects of the physical channel from the signal constellation design. The findings culminate in a definitive design directive: expanding the antenna's movement space is the primary means of improving performance, whereas simply increasing port density within a fixed area yields diminishing returns. This work offers not only an essential analytical tool but also foundational guidelines for the practical design and optimization of next-generation FAS.

\begin{appendices}

\section{Proof of Lemma 1}
The PDF of $\mathbf{h}=[{h}_1,h_2,\ldots,{h}_N]^T$ can be given by
\begin{equation}\label{hvect}
f(\mathbf{h})=\frac{1}{\pi^N\det(\mathbf{R})}\exp\left(-\mathbf{h}^H\mathbf{Ah}\right),
\end{equation}
where $\mathbf{A}=\mathbf{R}^{-1}$. Each element of $\mathbf{A}$ can be represented in polar form as
$
A_{ij} = |A_{ij}| e^{j \phi_{A_{{ij}}}}.
$

Considering (\ref{eq2}) and the variable transformation in (\ref{hvect}), the joint PDF of $g_1, \phi_1, \dots, g_N, \phi_N$ is given by
\begin{equation}
\begin{aligned}
&f(\mathbf{h})
% \\
% &
= \frac{\prod\nolimits_{n=1}^Ng_n\prod\nolimits_{n=1}^NH_n(g_n,\phi_n,\ldots,g_N,\phi_N)}{\pi^N\det(\mathbf{R})},
\end{aligned}
\end{equation}
where the term of $H_n(g_n,\phi_n,\ldots,g_N,\phi_N)$ denotes
\begin{equation}\label{Hnlar}
\begin{aligned}
&H_n(g_n,\phi_n,\ldots,g_N,\phi_N)\\=&\exp\left\{
-\mathbf{A}_{nn}g_n^2-2g_n\right.\\&\times\left.\left[
\sum\nolimits_{i=n+1}^Ng_i|\mathbf{A}_{in}|\cos(\phi_n-\phi_i+\phi_{\mathbf{A}_{in}})
\right]
\right\}.
\end{aligned}
\end{equation}
To streamline notation, we denote $H_n(g_n, \phi_n, \ldots, g_N, \phi_N)$ simply as $H_n$. Therefore, we have
\begin{equation}
\begin{aligned}
f(h_1, \ldots, h_N)=\frac{\prod_{n=1}^Ng_n}{\pi^N\det(\mathbf{R})}
{\int_{-\pi}^\pi\!\cdots\!\int_{-\pi}^\pi}\prod_{n=1}^NH_nd\phi_1\ldots d\phi_N.
\end{aligned}
\end{equation}
% The target PDF is found through a change of variables using the transformation $\gamma_{\rm FAS} = g_{\rm FAS}^2$.
To tackle the PDF of $g_{\rm FAS}$, we derive the CDF of $g_{\rm FAS}$ as
\begin{equation}
F(g_{\rm FAS})={\int_0^{g_{\rm FAS}}\cdots\int_0^{g_{\rm FAS}}}f(g_1,\ldots g_N)dg_1\ldots g_N.
\end{equation}
Based on this, the PDF of $g_{\rm FAS}$ can be evaluated as
\begin{equation}\label{fg_F}
\begin{aligned}
f (g_{\rm FAS})
%&=\frac{dF(g_{\rm FAS})}{dg_{\rm FAS}}\\
% =&N\underbrace{\int_0^{g_{\rm FAS}}\cdots\int_0^{g_{\rm FAS}}}_{N-1}f(g_1,\ldots,g_{\rm FAS})dg_1\ldots dg_{N-1}\\
=&\frac{Ng_{\rm FAS}}{\pi^N\det(\mathbf{R})}{\int_{-\pi}^{\pi}\cdots\int_{-\pi}^{\pi}}
H_NGd\phi_1\ldots\phi_N,
\end{aligned}
\end{equation}
where the term $G$ encapsulates the $N-1$ nested integrals over the channel gains $g_1,\ldots,g_{N-1}$ as
\begin{equation}
\begin{aligned}
G=&\int_0^{g_{\rm FAS}}H_{N-1}\int_0^{g_{\rm FAS}}H_{N-2}\\
&\cdots\left(\int_0^{g_{\rm FAS}}H_2\left(\int_0^{g_{\rm FAS}}H_1g_1dg_1\right)g_2dg_2\right).
\end{aligned}
\end{equation}
To evaluate this complex integral in $f(g_{\rm FAS
})$, we analyze its asymptotic behavior.
As such, we derive (\ref{Hnlar}) as
\begin{equation}
\int_0^{g_{\rm FAS}}H_n(g_n,\ldots,g_N)g_ndg_n=\frac{g_{\rm FAS}^2}{2}+o(g_{\rm FAS}^2),
\end{equation}
where
$H_N=1+o(1)$, for $g_{\rm FAS}\to 0$.
Starting from the innermost integral, we recursively solve the nested integrals of $G$. 
Applying the first approximation, we get
\begin{equation}
    \int_{0}^{g_{{\rm FAS}}}H_{1}g_{1}dg_{1} \approx \frac{g_{\rm FAS}^2}{2}.
\end{equation}
This result is then substituted into the next layer of the integral.
Since the term $\frac{g_{\rm FAS}^2}{2}$ is independent of the integration variable $g_2$, it can be factored out.
% Applying the same approximation to be remaining integral yields
% \begin{equation}
% \begin{aligned}
%       \int_{0}^{g_{{\rm FAS}}}H_{2}\left(\frac{g_{\rm FAS}^2}{2}\right)g_{2}dg_{2} &\approx \frac{g_{\rm FAS}^2}{2} \int_{0}^{g_{{\rm FAS}}}H_{2}g_{2}dg_{2} \approx \left(\frac{g_{\rm FAS}^2}{2}\right)^2.
% \end{aligned}
% \end{equation}
By extrapolating this logic to the entire set of $N-1$ nested integrals, we arrive at the approximation for
\begin{equation}
\begin{aligned}
    G &= \int_{0}^{g_{\rm FAS}}H_{N-1} \cdots \left(\int_{0}^{g_{\rm FAS}}H_{1}g_{1}dg_{1}\right) \cdots g_{N-1}da_{N-1} \\&\approx \left(\frac{g_{\rm FAS}^2}{2}\right)^{N-1}.
\end{aligned}
\end{equation}
With the asymptotic approximations for $G$ and $H_N$ established, we can finalize the derivation of $f(g_{\rm FAS})$. Substituting these results into (\ref{fg_F}) yields
\begin{equation}
\begin{aligned}
    f(g_{\rm FAS}) \approx &\frac{N g_{\rm FAS}}{\pi^{N} \det(\mathbf{R})} \\&\times{\int_{-\pi}^{\pi} \cdots \int_{-\pi}^{\pi}} (1) \left[ \left(\frac{g_{\rm FAS}^2}{2}\right)^{N-1} \right] d\phi_{1}\cdots d\phi_{N}.
\end{aligned}
\end{equation}
The integrand is now independent of the phase variables, allowing the $N$-fold integral to be readily evaluated as $(2\pi)^N$. Consequently, the expression for $f(g_{\rm FAS})$ simplifies to
\begin{equation}
\begin{aligned}
    f(g_{\rm FAS}) \approx \frac{N g_{\rm FAS}}{\pi^{N} \det(\mathbf{R})} \left(\frac{g_{\rm FAS}^2}{2}\right)^{N-1} (2\pi)^N = \frac{2N g_{\rm FAS}^{2N-1}}{\det(\mathbf{R})}.
\end{aligned}
\end{equation}
Having derived the asymptotic PDF of $g_F$, we proceed to determine the PDF of the effective SNR, $\gamma_{\rm FAS}$. This is accomplished by applying the change of variables technique to the transformation $\gamma_{\rm FAS} = g_{\rm FAS}^2$. The PDF of the transformed variable is given by
\begin{equation}
    f(\gamma_{\rm FAS}) = f(g_{\rm FAS}) \left| \frac{dg_{\rm FAS}}{d\gamma_{\rm FAS}} \right|_{g_{\rm FAS} = \sqrt{\gamma_{\rm FAS}}}.
\end{equation}
The Jacobian of this transformation is $|dg_F/d\gamma_{\rm FAS}| = 1/(2\sqrt{\gamma_{\rm FAS}})$ for $\gamma_{\rm FAS} > 0$. Substituting the expressions for $f(g_{\rm FAS})$ and the Jacobian, we arrive at the final result
\begin{equation}\label{frfR1}
\begin{aligned}
    f(\gamma_{\rm FAS}) 
    %&= \left[ \frac{2Ng_{\rm FAS}^{2N-1}}{\det(\mathbf{R})} \right]_{g_{\rm FAS} = \sqrt{\gamma_{\rm FAS}}} \cdot \frac{1}{2\sqrt{\gamma_{\rm FAS}}} \\
    % &= \frac{2N(\gamma_{\rm FAS})^{N - 1/2}}{\det(\mathbf{R})} \cdot \frac{1}{2\gamma_{\rm FAS}^{1/2}} \\
    &= \frac{N\gamma_{\rm FAS}^{N-1}}{\det(\mathbf{R})} = \frac{N\gamma_{\rm FAS}^{N-1}}{\det(\mathbf{J}) \prod_{n=1}^N \bar{\gamma}_n}.
    \end{aligned}
\end{equation}
This expression provides the sought-after PDF for $\gamma_{\rm FAS}$ and thus concludes the proof.

\end{appendices}


\begin{thebibliography}{99}

\bibitem{wong2022extrm}
K. Wong, K. Tong, Y. Chen, and Y. Zhang, ``Extra-large MIMO
enabling slow fluid antenna massive access for millimeterwave bands, \emph{Electron. Lett.}, vol. 58, no. 25, pp. 1016¨C1018, Dec. 2022.

\bibitem{new2024an}
X. Zhu, Q. Wu, W. Chen, Y. Liu, M. Jian, and D. B. D. Costa, ``Spatial Scattering Shift Keying for mmWave MIMO Systems," \emph{ IEEE Trans. Commun.}, early access, doi: 10.1109/TCOMM.2025.3597652.

\bibitem{zhu2025disc}
X. Zhu et al., ``Discrete reflecting beamforming optimization for RIS enhanced space shift keying MIMO systems," \emph{IEEE Wireless Commun. Lett.}, vol. 14, no. 8, pp. 2471-2475, Aug. 2025.

\bibitem{ris2025z}
X. Zhu et al., ``Performance analysis of RIS-aided double spatial scattering modulation for mmWave MIMO systems," \emph{IEEE Trans. Wireless Commun.}, vol. 23, no. 6, pp. 6139-6155, Jun. 2024.

% \bibitem{shoj2022mimo}
% A. Shojaeifard et al., ``MIMO evolution beyond 5G through reconfigurable intelligent surfaces and fluid antenna systems," \emph{Proc. IEEE}, vol. 110, no. 9, pp. 1244-1265, Sep. 2022.

% \bibitem{zhu2024per}
% X. Zhu et al., ``Performance analysis of RIS-aided double spatial scattering modulation for mmWave MIMO systems," \emph{IEEE Trans. Wireless Commun.}, vol. 23, no. 6, pp. 6139-6155, Jun. 2024.

% \bibitem{zhu2023risa}
% X. Zhu et al., ``RIS-aided spatial scattering modulation for mmWave MIMO transmissions," \emph{IEEE Trans. Commun.}, vol. 71, no. 12, pp. 7378-7392, Dec. 2023.

\bibitem{zhou2024movab}
Y. Zhou, W. Chen, Q. Wu, X. Zhu, and N. Cheng, ``Movable antenna empowered downlink NOMA systems: Power allocation and antenna position optimization," \emph{IEEE Wireless Commun. Lett.}, vol. 13, no. 10, pp. 2772-2776, Oct. 2024.

\bibitem{wong2022bruce}
K.-K. Wong, K.-F. Tong, Y. Shen, Y. Chen, and Y. Zhang,
``Bruce Lee-inspired fluid antenna system: Six research topics and
the potentials for 6G," \emph{Frontiers Commun. Netw.}, vol. 3, p. 5,
Mar. 2022.

\bibitem{wu2025scalable}
T. Wu, et al. ``Scalable FAS: A New Paradigm for Array Signal Processing." \emph{arXiv preprin}t arXiv:2508.10831, 2025.

\bibitem{yao2025fas}
J. Yao, T. Wu, L. Zhou, M. Jin, C. Huang, and C. Yuen, ``FAS vs. ARIS: Which is more important for FAS-ARIS communication systems?," \emph{IEEE Trans. Wireless Commun.}, early access doi: 10.1109/TWC.2025.3594617.

\bibitem{hong2025ge}
H. Hong, H. Xu and, X. Zhu, ``Geometric shaping non-uniform constellation design in fluid antenna system," in \emph{Proc. IEEE/CIC International Conference on Communications in China (ICCC Workshops)}, Shanghai, China, 2025, pp. 1-6

\bibitem{new2025at}
X. He, W. Chen, Q. Wu, X. Zhu, and N. Cheng, ``Movable antenna enhanced NOMA short-packet transmission," \emph{IEEE Commun. Lett.}, vol. 28, no. 9, pp. 2196-2200, Sept. 2024.



\bibitem{wong2021flns}
K. -K. Wong, A. Shojaeifard, K. -F. Tong, and Y. Zhang, ``Fluid antenna systems," \emph{IEEE Trans. Wireless Commun.}, vol. 20, no. 3, pp. 1950-1962, Mar. 2021.

\bibitem{wu2024flu}
T. Wu, et al. ``Fluid antenna systems enabling 6G: Principles, applications, and research directions." \emph{arXiv preprint arXiv:2412.03839}, 2024.

\bibitem{wong2020fles}
K. -K. Wong, W. K. New, X. Hao, K. -F. Tong, and C. -B. Chae, ``Fluid antenna system-Part I: Preliminaries," \emph{IEEE Commun. Lett.}, vol. 27, no. 8, pp. 1919-1923, Aug. 2023.

\bibitem{wong2023ii}
K. -K. Wong, K. -F. Tong, and C. -B. Chae, ``Fluid antenna system-Part II: Research opportunities," \emph{IEEE Commun. Lett.}, vol. 27, no. 8, pp. 1924-1928, Aug. 2023.

\bibitem{wong2023III}
K. -K. Wong, K. -F. Tong, and C. -B. Chae, ``Fluid antenna system-Part III: A new paradigm of distributed artificial scattering surfaces for massive connectivity," \emph{IEEE Commun. Lett.}, vol. 27, no. 8, pp. 1929-1933, Aug. 2023.

\bibitem{cha2024onms}
F. Rostami Ghadi, K. -K. Wong, W. K. New, H. Xu, R. Murch, and Y. Zhang, ``On performance of RIS-aided fluid antenna systems," \emph{IEEE Wireless Commun. Lett.}, vol. 13, no. 8, pp. 2175-2179, Aug. 2024.

\bibitem{wong2020pems}
K. K. Wong, A. Shojaeifard, K. -F. Tong, and Y. Zhang, ``Performance limits of fluid antenna systems," \emph{IEEE Commun. Lett.}, vol. 24, no. 11, pp. 2469-2472, Nov. 2020.

\bibitem{kwonglimi202}
K. -K. Wong and K. -F. Tong, ``Fluid antenna multiple access," \emph{IEEE Trans. Wireless Commun.}, vol. 21, no. 7, pp. 4801-4815, July 2022.

% \bibitem{wong2023oppf}
% K. -K. Wong, K. -F. Tong, Y. Chen, Y. Zhang, and C. -B. Chae, ``Opportunistic fluid antenna multiple access," \emph{IEEE Trans. Wireless Commun.}, vol. 22, no. 11, pp. 7819-7833, Nov. 2023.

\bibitem{won2023trsm}
K. -K. Wong, ``Transmitter CSI-free RIS-randomized CUMA for extreme massive connectivity'', \emph{IEEE Open J. Commun. Soc.}, vol. 5, pp. 6890-6902, Oct. 2024.

\bibitem{wang2003a}
Z. Wang and G. B. Giannakis, ``A simple and general parameterization quantifying performance in fading channels," \emph{IEEE Trans. Commun.}, vol. 51, no. 8, pp. 1389-1398, Aug. 2003.

% \bibitem{new2023flac}
% W. K. New, K. -K. Wong, H. Xu, K. -F. Tong, C. -B. Chae, and Y. Zhang, ``Fluid antenna system enhancing orthogonal and non-orthogonal multiple access," \emph{IEEE Commun. Lett.}, vol. 28, no. 1, pp. 218-222, Jan. 2024.

% \bibitem{zheng2025exp}
% J. Zheng, T. Wu, J. Yao, C. Yuen, Z. Ding, and F. Adachi, ``Exploring the impact of RIS on cooperative NOMA URLLC systems: A theoretical perspective," \emph{IEEE Trans. Wireless Commun.}, early access, doi: 10.1109/TWC.2025.3566602.

% \bibitem{he2024mov}
% X. He, W. Chen, Q. Wu, X. Zhu, and N. Cheng, ``Movable antenna enhanced NOMA short-packet transmission," \emph{IEEE Commun. Lett.}, vol. 28, no. 9, pp. 2196-2200, Sept. 2024.

% \bibitem{zheng2024fas}
% J. Zheng, T. Wu, X. Lai, C. Pan, M. Elkashlan, and K. -K. Wong, ``FAS-assisted NOMA short-packet communication systems," \emph{IEEE Trans. Veh. Technol.}, vol. 73, no. 7, pp. 10732-10737, Jul. 2024.

% \bibitem{cah2022pro}
% Z. Chai, K. -K. Wong, K. -F. Tong, Y. Chen, and Y. Zhang, ``Port selection for fluid antenna systems," \emph{IEEE Commun. Lett.}, vol. 26, no. 5, pp. 1180-1184, May 2022.

% \bibitem{kham2023anew}
% M. Khammassi, A. Kammoun, and M. -S. Alouini, ``A new analytical approximation of the fluid antenna system channel," \emph{IEEE Trans. Wireless Commun.}, vol. 22, no. 12, pp. 8843-8858, Dec. 2023.



% \bibitem{new2024flin}
% W. K. New, K. -K. Wong, H. Xu, K. -F. Tong, and C. -B. Chae, ``Fluid antenna system: New insights on outage probability and diversity gain," \emph{IEEE Trans. Wireless Commun.}, vol. 23, no. 1, pp. 128-140, Jan. 2024.









% \bibitem{ser1998on}
% S. Serra, ``On the extreme eigenvalues of Hermitian (block) Toeplitz
% matrices," \emph{Linear Algebra Appl.}, vol. 270, nos. 1--3, pp. 109--129, 1998.

% \bibitem{zhu2025trans}
% X. Zhu, Q. Wu, and W. Chen, ``Transmissive RIS transmitter enabled spatial modulation MIMO systems," \emph{IEEE J. Sel. Areas Commun.}, vol. 43, no. 3, pp. 899-911, Mar. 2025.



% \bibitem{zhu2024robust}
% X. Zhu, W. Chen, Q. Wu, W. Fang, C. Huang, and J. Li, ``Robust analysis of full-duplex two-way space shift keying with RIS systems," \emph{IEEE Trans. Commun.}, vol. 72, no. 11, pp. 7233-7249, Nov. 2024.




\bibitem{zhu2025toward}
X. Zhu, Q. Wu, W. Chen, Z. Zhang, X. Bai, and X. Guan, ``Towards spatial scattering modulation: Detector design and error probability analysis," \emph{IEEE Trans. Wireless Commun.}, early access, doi: 10.1109/TWC.2025.3567484.

\bibitem{zhu2024on}
X. Zhu, Q. Wu, and W. Chen, ``On the performance of RIS-aided spatial modulation for downlink transmission," \emph{IEEE Trans. Wireless Commun.}, vol. 23, no. 11, pp. 16203-16217, Nov. 2024.

% \bibitem{zhu2023ris}
% X. Zhu et al., ``RIS-assisted full-duplex space shift keying: System scheme and performance analysis," \emph{IEEE Trans. Green Commun. Netw.}, vol. 7, no. 4, pp. 1981-1995, Dec. 2023.



\end{thebibliography}
\end{document}